# Spectral characterization of a frequency comb based on cascaded quadratic nonlinearities inside an optical parametric oscillator

Ville Ulvila,[1] C. R. Phillips,[2] Lauri Halonen,[1] and Markku Vainio[1,3,*]

[1]*Laboratory of Physical Chemistry, Department of Chemistry, P.O. Box 55, 00014 University of Helsinki, Finland*
[2]*Department of Physics, Institute of Quantum Electronics, ETH Zürich, 8093 Zürich, Switzerland*
[3]*VTT Technical Research Centre of Finland Ltd, Centre for Metrology MIKES, P.O. Box 1000, FI-02044 VTT, Finland*
[*]*markku.vainio@helsinki.fi*

**We present an experimental study of optical frequency comb generation based on cascaded quadratic nonlinearities inside a continuous-wave-pumped optical parametric oscillator. We demonstrate comb states which produce narrow-linewidth intermode beat note signals, and we verify the mode spacing uniformity of the comb at the Hz level. We also show that spectral quality of the comb can be improved by modulating the parametric gain at a frequency that corresponds to the comb mode spacing. We have reached a high average output power of over 4 W in the near-infrared region, at ~2 μm.**

## I. INTRODUCTION

In the last two decades, several novel methods for optical frequency comb generation have been reported in the literature [1-3]. In 2007, the Kerr comb, a new method based on continuous-wave-pumped (CW-pumped) optical microresonators, was discovered [4]. Kerr comb generation utilizes the high quality factor of microresonators which enhances the cubic ($\chi^{(3)}$) nonlinearity of the material and allows broadband frequency comb generation by four-wave-mixing [5-8]. More recently, we have demonstrated that analogous frequency comb generation with a CW-pump laser is also possible in a $\chi^{(2)}$ –system, which mimics the cubic nonlinearity [9, 10] via cascaded quadratic ($\chi^{(2)}$) nonlinearities (CQN) [11], such as phase-mismatched second harmonic generation (SHG). This scheme is here referred to as a CQN comb generation. The absolute value of the effective nonlinear refractive index $n_2$ arising from the cascaded quadratic nonlinearities can be several orders of magnitude larger than that of the inherent $n_2$ of the material, and the sign of the effective $n_2$ can be changed by varying the phase mismatch of the SHG interaction. These properties lead to significant advantages that complement the microresonator Kerr combs: The CQN concept also makes efficient CW-pumped comb generation possible in bulk resonators that have modest finesse. The bulk resonator can be used to produce a versatile high-power frequency comb with a mode spacing of ~100 MHz to 10 GHz. This is desirable for many applications like trace gas spectroscopy [3, 12-19]. In addition to yielding a high power per comb line, the operating wavelength of the CQN comb is widely tunable, only limited by the transparency of the crystal material and not by the available laser wavelength [10]. The cascaded quadratic nonlinearities have also proven useful in many other applications, such as supercontinuum generation [20, 21], mode-locking of femtosecond lasers [22-26], pulse compression and optical soliton generation [27-33].

In this work, we have used a CW-pumped singly resonant optical parametric oscillator (OPO), which incorporates an additional nonlinear crystal for frequency comb generation by the cascaded quadratic nonlinearities. As well as showing great technological promise, this CQN concept presents us with a new physical platform for comb generation. While cascaded quadratic nonlinearities are Kerr-like, this approximation breaks down close to phase-matching, which will have a significant influence on how strong the effective nonlinearity can be made inside the OPO cavity. The physical processes underlying comb generation inside a singly-resonant OPO differ significantly from parametric comb formation in conventional $\chi^{(3)}$ based systems. For example, in $\chi^{(3)}$-systems, modulation instabilities (MIs) play an important role in generating new modes from an initially CW laser [34-36], and MIs are also connected with periodic solutions associated with a comb of sidebands [37]. In a singly-resonant OPO, the MI properties fundamentally differ from $\chi^{(3)}$-systems and can have a significant effect on OPO operation and stability [38]. In a CQN based comb inside a singly-resonant OPO, we combine the two types of systems (Kerr-like nonlinearity inside singly-resonant OPO), and can therefore expect a rich variety of physical processes to be accessible for frequency comb generation.

In this article, we demonstrate that an optical frequency comb generated by the cascaded quadratic nonlinearities can produce narrow-linewidth intermode beat note signals similar to those observed with Kerr combs, and that spectral quality of the comb can be further improved by modulating the parametric gain via pump laser intensity modulation. The most important result of the paper is the first rigorous verification of the mode spacing uniformity of a CQN comb. This is an important step towards practical applications, which often require a strictly uniform comb structure [1-3].

## II. PRINCIPLE OF OPERATION AND EXPERIMENTAL SETUP

The principle of comb formation in our system is illustrated in Fig. 1. The use of CW pumping without the need for active modulation makes our frequency comb generation method unique, when comparing to more traditional intracavity modulated OPOs [39-42] and synchronously pumped OPOs [43-50]. It is also possible to generate frequency combs based on cascaded quadratic nonlinearities with a simple optical setup based on a cavity-enhanced SHG, as has recently been demonstrated by Ricciardi *et al.* [51]. However, a singly-resonant OPO based approach is not limited by the wavelength of the pump laser, and can support ultra-broadband tuning from visible to the important mid-infrared region [10, 52]. With our CW-pumped OPO, the pump is non-resonant and does not need to be locked to the cavity modes, since the cascading quadratic nonlinearities are initiated by the OPO's signal wave which is automatically resonating inside the cavity once the OPO is running. Moreover, the singly-resonant OPO makes it possible to achieve a high average optical output power of several watts.

Our experimental setup is shown in Fig. 2. The first nonlinear crystal is a 50 mm long periodically poled MgO:LiNbO$_3$-crystal (PPLN 1, HC Photonics) with a fan-out quasi-phasematching grating. The poling period can be tuned continuously between 26.5 – 32.5 μm by translating the crystal. This crystal is responsible for the OPO gain, i.e., it converts pump photons to signal and idler photons. The pump laser is a high-power Yb-fiber amplifier ($\lambda_p$ = 1064 nm, IPG Photonics YAR-20K-1064-LP-SF), seeded with a narrow-linewidth distributed feedback diode laser. Up to 18 W of pump power is delivered to PPLN 1. The OPO oscillation threshold varies depending on the exact operation parameters, but is typically achieved at pump powers between 3 and 9 W.

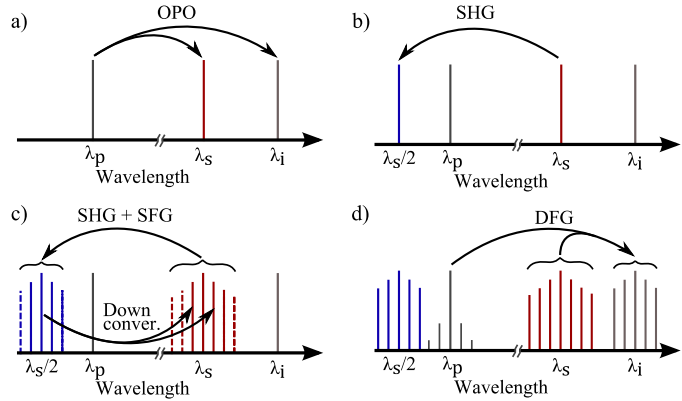

Fig. 1. (Color online) CQN comb generation in a singly-resonant OPO. a) Signal and idler photons are generated from the pump photons ($1/\lambda_p = 1/\lambda_s + 1/\lambda_i$, where $\lambda_p$, $\lambda_s$, and $\lambda_i$ are the wavelengths of the pump, signal, and idler beams, respectively). The signal wave resonates in the OPO cavity. In b) and c), cascaded quadratic nonlinearities lead to comb formation (SFG = sum frequency generation). d) The comb structure is transferred to the idler wave by difference frequency generation (DFG). Back conversion of the signal and idler combs also creates a weak comb structure in the depleted pump wave.

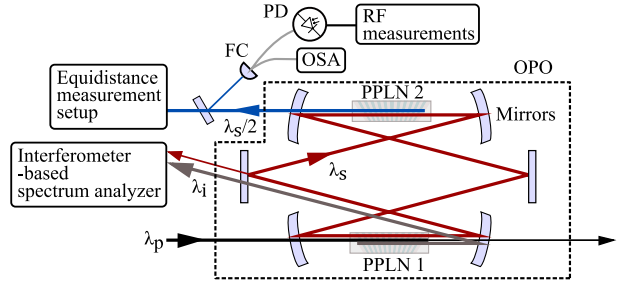

Fig. 2. (Color online) A schematic overview of the OPO and measurement setups. OSA: optical spectrum analyzer, FC: fiber coupler, PD: photodetector, and RF: radio frequency. An offset to the optical beams involved has been drawn just for clarity.

The second nonlinear crystal (PPLN 2), which produces the cascading quadratic effect for the resonant signal wave of the OPO, is identical to PPLN 1. The total average optical power at the second harmonic (SH) frequency of the signal wave is 1 - 100 mW, depending on the amount of phase mismatch for the SHG of the signal wave in PPLN 2. The OPO bow-tie ring cavity is designed such that the waists of the resonant signal beam are in the middle of the crystals. The $1/e^2$-radii of the waists are 62 µm. The concave mirrors of the cavity have high reflectivity for the 1.5 – 2.5 µm range but the plane mirrors only for the signal wavelengths 1.73 – 2.05 µm. This gives an idler wavelength tuning range of approximately 2.2 – 2.8 µm. The free spectral range (FSR) of the OPO cavity is ~207 MHz, which also determines the mode spacing of the frequency comb. The OPO cavity length was not actively stabilized in the experiments reported here.

An interferometer-based spectrum analyzer (EXFO WA-1500-NIR/IR-89 + EXFO WA-650) was used to record the optical spectra of the signal and idler combs at wavelengths longer than 1.75 µm. Below that, a grating-based spectrum analyzer (OSA, Ando AQ-6315E) was used to record the optical spectra of the signal SH comb and externally frequency-doubled idler comb. A fast InGaAs-photodetector (Thorlabs DET01CFC, bandwidth 1.2 GHz) connected to an RF spectrum analyzer (Agilent 4395A, bandwidth 500 MHz) was used to measure the RF spectra of the signal SH comb and frequency-doubled idler comb. Lack of a fast mid-infrared photodetector prohibited direct RF measurements of the signal and idler combs.

## III. RESULTS

We operated both crystals at 60 °C temperature, and used poling period $\Lambda_1 \sim 32.2$ µm for PPLN 1, thus generating $\lambda_s \sim 2036$ nm and $\lambda_i \sim 2228$ nm OPO wavelengths. This signal wavelength was chosen because it is close to the calculated zero group delay dispersion (GDD) wavelength of the OPO cavity (Fig. 3). Another region of small cavity GDD is close to the zero dispersion wavelength of the PPLN-crystals (~1.92 µm), but operating a bulk OPO in this region is challenging because of strong absorption due to the ambient air. We observed comb formation over a large poling period range of PPLN 2, $\Lambda_2 \sim 29 - 32$ µm. This corresponds to a wave-vector mismatch $\Delta k \sim -12$ to $+8$ mm$^{-1}$ [9]. Here, $\Delta k = k_{2\omega} - 2k_\omega - 2\pi/\Lambda_2$, where $k_\omega$ and $k_{2\omega}$ are the wave-vectors of the signal and signal SH waves, respectively. With $\Lambda_2 \sim 31.2$ µm, 18 W of pump power, and pump depletion of ~50 %, the obtained average output powers of the signal and idler combs are high, 0.9 W and 4.2 W, respectively. The high output power (0.9 W) at the signal wavelength arises because this wavelength is close to the edge of the high reflectivity region of the plane mirrors, which therefore act as output couplers for the signal comb.

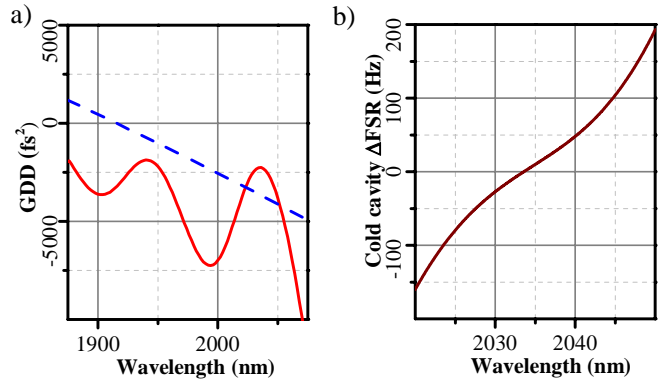

Fig. 3. (Color online) a) The dashed blue line shows the calculated GDD caused by PPLN-crystals for a single cavity round trip. The solid red line shows the calculated total GDD caused by the OPO cavity mirrors (Layertec GmbH) and the PPLN-crystals. b) Calculated change of the cold cavity free spectral range (FSR) of the OPO when taking into account cavity GDD (The reference wavelength 2033.6 nm, where FSR = 206.65 MHz, was arbitrarily chosen). For the refractive indices of the PPLN-crystals, Gayer *et al.* [53] data were used.

Recent work with Kerr combs has revealed two distinctive scenarios of comb formation [54]. In the first one, the comb generation starts from the modes adjacent to the pump mode. In our case this "pump mode" would correspond to the OPO signal mode that first arises as the OPO exceeds the threshold. In the other scenario, the comb formation starts by parametric generation of side bands, which are located at multiple FSRs from the initial pump mode [54-56]. We observe similar behavior with the CQN comb. In our previous publications we have concentrated on the first scenario, where the comb formation starts from the adjacent modes [9, 10]. However, side band generation at multiple FSRs has also been observed [51, 57], and can be qualitatively explained by theoretical models that have recently been developed to describe such systems [51, 58].

In most cases, our CQN comb starts to grow adjacent to the "pump mode", resulting in an optical envelope spectrum that is broad and continuous [9, 10]. The corresponding intermode beat note is also broad, the typical full-width at half-maximum (FWHM) being approximately 1 MHz. However, with certain poling period (phase mismatch) values, the beat note narrows down to a sharp peak on top of a pedestal. For example, with $\Lambda_2 \sim 31.2$ μm ($\Delta k \sim + 3$ mm$^{-1}$), the FWHM linewidth of an intermode beat note is <5 kHz (Fig. 4).

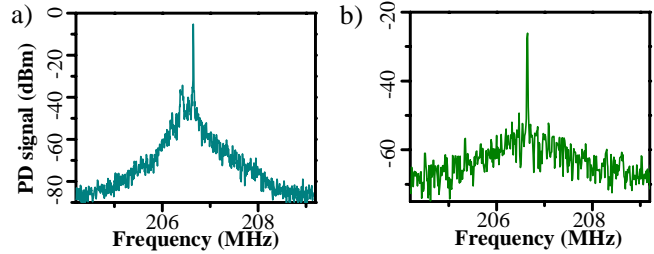

Fig. 4. (Color online) a) An RF spectrum of the signal SH comb. Resolution bandwidth (RBW): 3 kHz, sweep time: 271.4 ms. b) An RF spectrum of the externally frequency-doubled idler comb. RBW: 10 kHz, sweep time: 274.1 ms.

The regime of narrow intermode beat notes is typically associated with an optical spectrum that consists of several distinct side bands that are separated by multiple FSRs from the initially resonating OPO signal mode [Fig. 5(a) and 5(b)]. With the experimental setup reported here, the side band separation is typically 71 GHz. However, this separation can be changed, for example, to 71 GHz/2 = 35.5 GHz by slightly changing the OPO parameters or by gently disturbing its operation. While the dynamic range of our interferometer-based spectrum analyzer is insufficient to observe the spectral features between the strong peaks, the OSA measurements of the frequency-doubled signal/idler spectra [Fig. 5(c) and 5(d)] suggest that these sub combs overlap with each other. It is also possible that the OPO signal and idler spectra [Figs. 5(a) and (b)] consist of separate combs (side bands) that do not overlap with each other but still produce continuous SH combs [Fig. 5(c) and 5(d)] due to parametric mixing. In any case, in this regime, the RF spectrum measured of the signal SH comb often contains weak side modes [Fig. 4(a)], which could imply the presence of sub combs with different offset frequencies [54].

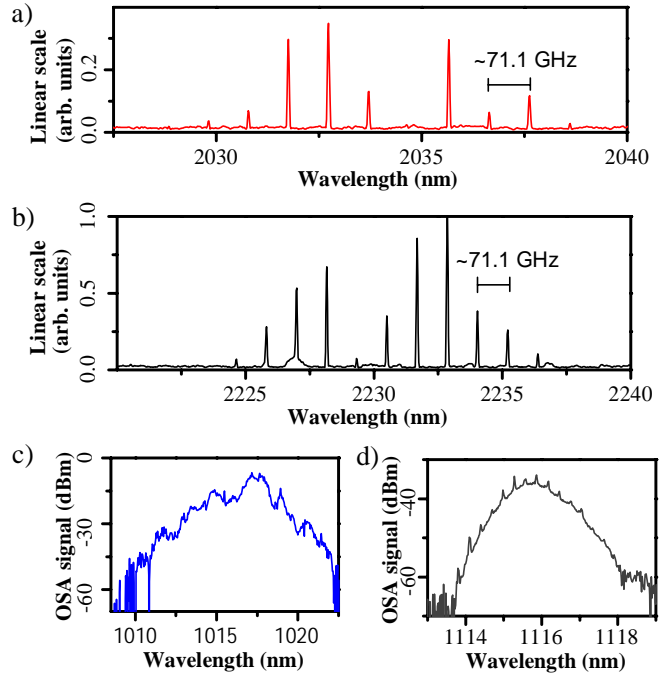

Fig. 5. (Color online) The optical envelope spectra of a) the signal and b) the idler comb, showing several strong side bands. The spectra shown in figures a and b were recorded with an interferometer-based spectrum analyzer. The optical envelope spectra of c) the signal SH and d) the externally frequency doubled idler comb. The spectra shown in figures c and d were recorded with an OSA.

In order to improve the spectral quality of the comb, we applied a parametric seeding technique similar to that used with microresonator Kerr combs [59, 60]. In our approach, we modulated the power of the pump laser system via the current of the seed laser. The amplified pump power follows this modulation if the modulation frequency is high enough. This intensity modulation creates side bands around the pump wavelength, which are then also generated around the signal (and idler) wavelength via parametric gain. When the modulation frequency is tuned to the intermode beat note frequency (comb mode spacing), it seeds the adjacent comb modes and reduces the phase noise significantly, as is demonstrated in Fig 6. With a modulation depth of 8%, which is the maximum modulation depth tolerated by the pump laser amplifier, the RF beat side modes disappear and the noise pedestal significantly drops by more than 20 dB around the beat note frequency. Approximately 99% and 82% of the power is concentrated in the sharp center peak of the intermode beat note [Fig. 6(a)] in the seeded and unseeded case, respectively. The effect of seeding is also seen in the time domain, where the CW background and amplitude variation of the intermode beat signal decrease [Fig. 6(b)]. The locking range is estimated to be ~1.5 kHz, while the comb mode spacing typically drifts by ~1 kHz, or $5\times10^{-6}$ relative to the OPO cavity FSR, in 10 minutes with the unstabilized cavity used in these experiments. This parametric seeding technique, which is essentially an active modulation method, offers a simple way to improve spectral quality of the CQN comb without any additional optical components, such as acousto- or electro-optic modulators. Passive methods to improve the quality of the comb by selecting suitable operating parameters or configuration of the OPO, e.g. by optimizing the dispersion of the OPO cavity, system is one of the most important topics of the future research.

Many applications of optical frequency combs, such as frequency metrology and dual-comb spectroscopy, require the comb to be precisely equidistant and the mode spacing to be adjustable. We have previously experimentally showed that the CQN comb mode spacing can be controlled by varying the cavity length, allowing for stabilization of the comb to an optical or radiofrequency reference [10]. Here, we report a major advancement in verifying the uniformity of the CQN comb mode spacing. The measurements were performed using a simple method that does not require a reference frequency comb [56]. This method is based on comparison of the mode spacing at two different parts of the comb spectrum, as explained in detail in Appendix. In brief, two different parts of the signal SH comb spectrum are separated by a diffraction grating. These are then coupled to their own photodetectors, which create independent intermode beat frequencies. One of the frequencies is shifted by 10 MHz by mixing it with a reference signal. Finally, the shifted beat frequency is mixed with the other beat frequency to create a signal at 10 MHz. Any deviation from 10 MHz would mean that the comb mode spacing is not uniform.

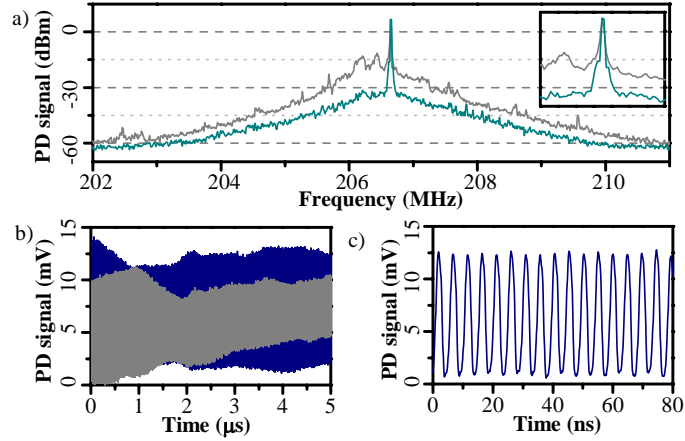

Fig. 6. (Color online) Parametric seeding experiments. a) RF spectrum of the CQN comb. The dark cyan line: Pump laser intensity modulation depth 8% and the modulation frequency tuned to the comb intermode beat note frequency. The gray line: Without the pump intensity modulation. Both spectra averaged from 16 consecutive scans. RBW: 10 kHz, sweep time: 274.1 ms. Inset: Close-up of the peak center (span 0.7 MHz). b) Time-domain photodetector signal. The blue line: Modulation depth 8 %. The gray line: Without the pump intensity modulation. c) A close-up of the blue line of Fig. 6(b).

Figure 7(a) shows an example of mode-spacing uniformity measurement, which was done by comparing two parts of the signal SH comb at 1014 nm and 1018 nm. The final-stage signal measured with an RF spectrum analyzer shows a beat note exactly at 10 MHz, as expected. A series of RF spectra from the measurement is plotted in Fig. 7(b), with a mean center value of 10 MHz and a standard deviation of 0.2 Hz. The measurements were repeated several times, such that different parts of the signal SH comb were compared. The signal SH comb was also compared against the frequency-doubled idler comb. In all cases the frequency of the final-stage signal was measured to be exactly 10 MHz. We note that these measurements from the signal SH comb and frequency-doubled idler comb do not necessarily imply alone that the signal comb mode spacing should also be uniform, but due to the law of conservation of energy there is no reason to think otherwise, although different combs can have different phase-noise properties [51] and different offset frequencies. In any case, these measurements strongly indicate that the CQN comb mode spacing is uniform within the measurement resolution (1 Hz) of the RF spectrum analyzer. This experimental uncertainty is orders of magnitude better than in the previous mode spacing measurements of CQN combs [9, 10, 51] and significantly smaller than the change of the OPO cold cavity FSR, which is approximately 80 Hz within the signal comb span (Fig. 3). In the case where the intermode beat note is broad (FWHM ~ 1 MHz) [9, 10], the resulting final-stage signal in the comb mode uniformity measurement is also broad, with a width comparable to that of the original intermode beat note signal.

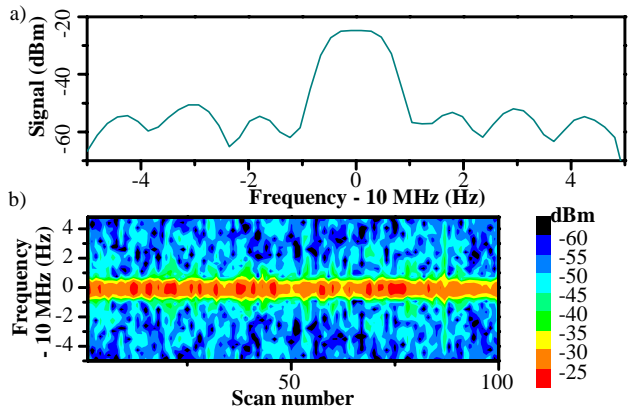

Fig. 7. (Color online) Mode-spacing uniformity measurement. a) An example of an RF beat note of the final stage of the measurement. The signal center is at 10 MHz, FWHM = 0.88 Hz. RBW: 1 Hz, and sweep time: 5.471 s. b) A contour plot that shows 100 consecutive scans with the RF spectrum analyzer.

## IV. CONCLUSION

In conclusion, we have investigated optical frequency comb generation by cascaded quadratic nonlinearities at wavelengths longer than 2 µm. We have obtained an average output power of over 4 W by using an implementation based on a CW-pumped singly-resonant OPO. We have, for the first time, confirmed that the produced spectrum is indeed a frequency comb and not just a collection of modes oscillating at nonuniform cold cavity eigenfrequencies in the presence of dispersion. This was done for a state where the comb formation starts from multiple mode spacing. We have demonstrated that spectral quality of such a comb can be improved by parametric seeding.

Our new results together with previously published results (e.g., methods to control the offset frequency and the mode spacing of the comb [10]) open up a simple way to generate frequency combs in a wide range of wavelength regions; in principle the CQN comb generation is limited only by the transparency of the nonlinear material used. For example, phase-matched DFG between solitary comb lines of the resonating signal comb could inherently produce comb lines in the far-infrared region (THz-region) in the PPLN-crystals used in these experiments [61]. Moreover, the CQN comb may prove to be a versatile platform to study the properties of the Kerr combs. In particular, the CQN scheme makes it possible to vary the sign and magnitude of the effective nonlinear refractive index, while in Kerr combs these are fixed material parameters.

# ACKNOWLEDGEMENTS

We are grateful to the University of Helsinki, the Academy of Finland, and the Magnus Ehrnrooth Foundation for funding the research reported in this contribution.

### Appendix: Uniformity measurement setup

The measurement setup used to confirm the uniformity of the CQN comb line spacing is based on a method which does not rely on a reference comb [56]. In this method, the line spacing of two different parts of the comb spectrum are directly compared with each other.

The signal second harmonic (SH) comb from the CQN comb generator, which is based on a continuous-wave-pumped optical parametric oscillator (OPO), is aligned to a diffraction grating. The grating selects two different wavelength regions of the comb. They are separately coupled to optical fibers, which then lead the light to independent photodetectors (PD). This is shown in Fig. A1, where part A of the signal SH comb is coupled to PD A and part B to PD B. The photodetectors create electronic signals which contain intermode beat frequencies corresponding to the line spacings of the different parts of the comb. Here, these beat frequencies are named $f_A$ and $f_B$. Both beat-frequency signals are band-bass filtered and amplified, see Fig. A1.

A reference signal $f_{ref}$ = 10 MHz is mixed with the beat frequency $f_A$ (in Mixer A), in order to transfer the uniformity comparison from DC to a frequency that can easily be measured by a radio frequency (RF) spectrum analyzer or a frequency counter. The reference signal is produced by an internal oscillator of the same RF spectrum analyzer that is used in the mode-spacing uniformity measurements.

After Mixer A, signal $f_A \pm f_{ref}$ is filtered so that the original reference frequency cannot leak to the next mixer (Mixer B). In Mixer B, $f_B$ is mixed with the $f_A \pm f_{ref}$ signal, creating the final signal $f_{ed} = |f_A \pm f_{ref} \pm f_B|$. If the intermode beat frequencies $f_A$ and $f_B$ are exactly the same, $f_{ref}$ should be reproduced in the $f_{ed}$ signal. This can be checked by monitoring $f_{ed}$ with an RF spectrum analyzer. An example of RF spectrum of $f_{ed}$ is shown in Fig. 7(a). If the frequencies are not the same, i.e., if the comb is non-uniform, one observes two separate peaks at $f_{ref} \pm |f_A - f_B|$.

The accuracy of the measurement is limited by the resolution of our RF spectrum analyzer, which is 1 Hz. This was verified by feeding the analyzer's own reference signal to the analyzer. A better resolution could be obtained with a frequency counter [32]. However, the strong amplitude variation of the comb intermode beat note, which can exceed 10 dB, prevents reliable use of a frequency counter in our case.

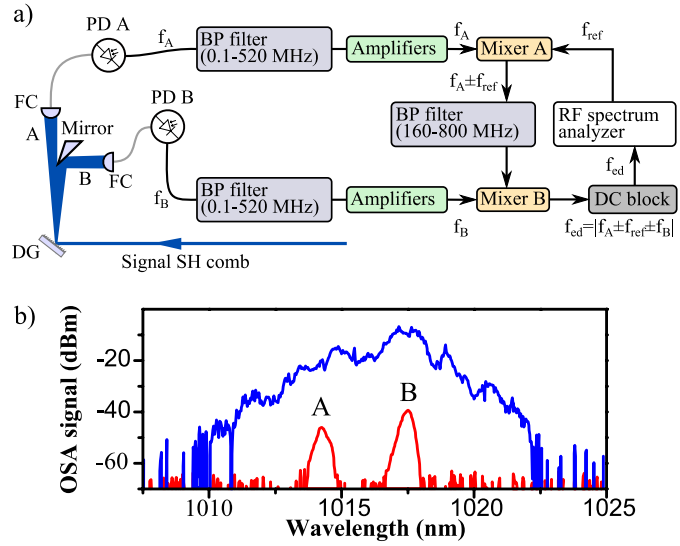

Fig. A1. (Color online) a) The mode-spacing uniformity measurement setup. DG: diffraction grating, FC: fiber coupler, PD: photodetector, BP: band pass, DC: direct current, RF: radio frequency. b) An example of the signal SH comb optical spectrum (blue line). The red line shows two parts of the comb spectrum that are extracted with the DG to separate photodetectors. The measurement was repeated several times with different parts of the signal SH comb and also between the signal SH comb and externally frequency doubled idler comb.